# Time and length scales of autocrine signals in three dimensions

Mathieu Coppey*, Alexander M. Berezhkovskii# Stuart C. Sealfon§, Stanislav Y. Shvartsman*

* Department of Chemical Engineering and Lewis-Sigler Institute for Integrative Genomics, Princeton University, New Jersey ; # Mathematical and Statistical Computing Laboratory, Division of Computational Bioscience, Center for Information Technology, National Institutes of Health, Bethesda, MD 20892 §; Department of Neurology and Center for Translational Systems Biology, Mount Sinai School of Medicine, New York.

ABSTRACT:   A model of autocrine signaling in cultures of suspended cells is developed on the basis of the effective medium approximation. The fraction of autocrine ligands, the mean and distribution of distances traveled by paracrine ligands before binding, as well as the mean and distribution of the ligand lifetime are derived. Interferon signaling by dendritic immune cells is considered as an illustration.

Address reprint requests and inquiries to Stanislav Y. Shvartsman, email: stas@princeton.edu

## INTRODUCTION

Autocrine loops can control the self-renewal and differentiation of stem cells (1), establish the spatial patterns of cell fates in development (2), enable local tissue repair (3), and protect cells from a variety of stresses (4). In addition to their ubiquitous role in tissues, autocrine loops can operate in experiments with cells in culture (5-9). These experiments can be done in one of the two formats. In the case of experiments with cultures of adherent cells, the cells are distributed in two dimensions and secreted ligands are diffusing in the overlaying culture medium (10,11). In the second format, cells are suspended in the three-dimensional medium (12-16).

Independently of the experimental format, one is frequently interested in the following properties of ligand trajectories (5,6,8,9,17). First, it is important to determine the probability that a ligand trajectory, initiated at the cell surface, is recaptured by the same ("parent") cell. This probability is denoted by $P_{auto}$. Clearly, $P_{para} \equiv 1 - P_{auto}$, is the fraction of the ligands that bind to the cell's neighbors. In this paper, the ligands recaptured by the parent cell are called "autocrine", while the ones captured by the cell's neighbors are called "paracrine". Next, it is important to determine the distribution of lifetimes for autocrine and paracrine ligands. The corresponding probability densities are denoted by $\varphi_{auto}(t)$ and $\varphi_{para}(t)$. Based on these probability densities one can define the average lifetimes of autocrine and paracrine ligands, $<t_{auto/para}>$, which provide the natural time scales for autocrine and paracrine signals. The length scale of paracrine signals can be obtained from the probability density of the ligand trapping points, $p_{para}(r)$, and its first moment, $\langle r_{para} \rangle$.

While it is difficult to measure these properties of ligand trajectories, they might be predicted on the basis of biophysical models or extracted from measurements of cellular responses (18,9). One of the goals of modeling is to connect these experimentally inaccessible properties of autocrine systems to the properties of individual cells, such as the levels of receptor expression, and parameters of the culture, such cell densities and medium volumes (17).

Recently, we have developed models for autocrine signaling in experiments with epithelial layers and cultures of adherent cells (10,19-22). We have shown that this problem is effectively one-dimensional and can be efficiently handled using a boundary homogenization approach, whereby the heterogeneous surface of the tissue culture plate is approximated by a partially absorbing boundary condition, which depends on the properties of individual cells and the cell surface fraction (19,23,10). In this problem, the height of the liquid medium which covers the layer of adherent cells is an important parameter that controls the spatial and temporal characteristics of ligand trajectories (19,24). The geometry of the cell communication in cultures of suspended cells is completely different; hence, a new formalism is required for its analysis. The three-dimensional format is frequently encountered both in experiments with suspended cell cultures and in *in-vivo* (12-16,25).

Our analysis is motivated by the characterization of the autocrine and paracrine signals in cultures of dendritic cells (26). In response to viral infection, these cells start secreting IFNβ, which can affect both the parent cells and their uninfected neighbors. The secretion of the virus-induced IFNβ is essential for the maturation of dendritic cells. In this context, it is important to determine what fraction of secreted IFNβ is recaptured by the ligand-secreting cells, to estimate how long do these ligands spend in the medium, and to establish their spatial range. In this paper we show how to derive the analytical expressions for all of these important properties.

Our approach is based on the effective medium approximation: the three-dimensional heterogeneous medium with randomly distributed cells is replaced by a uniformly absorbing medium characterized by a volumetric trapping rate constant, which depends on the cell density, the ligand diffusivity, and the properties of individual cells (their size, the level of receptor expression, and the rate constant of ligand-receptor binding), Figure 1. The paper is organized as follows. In the next section, we present the main results for the statistical properties of autocrine and paracrine trajectories. Next, we outline the main steps for their derivation. Then, we illustrate the application of these results to specific experiments with cultured dendritic cells. Finally, we conclude with the discussion of our results and outline the steps for their incorporation into more complex cellular and biochemical models of autocrine signaling.

## MODEL FORMULATION AND SUMMARY OF ANALYTICAL RESULTS

Consider spherical cells of radius $R$, which are uniformly distributed in the three-dimensional medium. The concentration of cells is denoted by $c$. Each cell has a fixed number of receptors, which is denoted by $N_R$. Ligands, diffusing in the medium with diffusivity $D$, bind to receptors with the rate constant $k_{on}$. Ligand binding to individual cells is characterized by an effective surface trapping rate $\kappa$ (27,28):

$$\kappa = \frac{k_{on} N_R}{4\pi R^2} \qquad 2.1$$

Using effective medium approximation, we introduce the volumetric rate constant, $k_b$, which describes trapping of a ligand diffusing in the culture of suspended cells (29,4):

$$k_b = \frac{k_{Sm} k_{on} N_R}{k_{Sm} + k_{on} N_R} c = \frac{k_{Sm} R\kappa}{D + R\kappa} c \qquad 2.2$$

where $k_{Sm} = 4\pi D R$ is the Smoluchowski rate constant (30).

Our results can be most conveniently expressed in terms of the dimensionless surface trapping rate, $\tilde{\kappa}$:

$$\tilde{\kappa} \equiv \frac{\kappa R}{D} = \frac{k_{on} N_R}{4\pi D R} \qquad 2.3$$

which is the ratio of the trapping probability to the escape probability for a ligand secreted by an isolated cell (4,28). This leads to the following expression for $k_b$:

$$k_b = c \frac{4\pi R^2 \kappa}{1+\tilde{\kappa}} = c \frac{k_{on} N_R}{1+\tilde{\kappa}} \qquad 2.4$$

The dimensionless form of $\tilde{k}_b$, given by the product of $k_b$ and the characteristic diffusion time, $R^2/D$, can be written in terms of $\tilde{\kappa}$ and the cell volume fraction, $\omega = (4/3)\pi R^3 c$:

$$\tilde{k}_b \equiv \frac{k_b R^2}{D} = \frac{3\omega \tilde{\kappa}}{1+\tilde{\kappa}} \qquad 2.5$$

In the rest of this section we present our main results; their derivation is given in the next section.

The survival probability of the ligand released at $t=0$, $S(t)$, is given by

$$S(t) = \left\{ 1 - \frac{\tilde{\kappa}}{1+\tilde{\kappa}} \left[ 1 - e^{(1+\tilde{\kappa})^2 k_b t / \tilde{k}_b} \operatorname{erfc}\left( (1+\tilde{\kappa}) \sqrt{k_b t / \tilde{k}_b} \right) \right] \right\} e^{-k_b t} \qquad 2.6$$

Where $\operatorname{erfc}(z)$ is the complementary error function (31). The mean lifetime of the ligand, $\langle t \rangle$, is given by

$$\langle t \rangle = \frac{1+\sqrt{\tilde{k}_b}}{k_b \left(1+\tilde{\kappa}+\sqrt{\tilde{k}_b}\right)} = \frac{(R^2/D)\left(1+\sqrt{\tilde{k}_b}\right)}{\tilde{k}_b \left(1+\tilde{\kappa}+\sqrt{\tilde{k}_b}\right)} \qquad 2.7$$

The second expression provides the relation between the average lifetime $\langle t \rangle$ and the characteristic diffusion time, $R^2/D$.

The probability density for the distribution of the ligand trapping points, $p(r)$, has the following form:

$$p(r) = \frac{\tilde{\kappa}\delta(r-R) + \tilde{k}_b \exp\left[-(r/R-1)\tilde{k}_b^{1/2}\right] H(r-R)/r}{4\pi R^2 \left(1 + \tilde{\kappa} + \tilde{k}_b^{1/2}\right)} \qquad 2.8$$

where $r$ is the distance between the trapping point and the center of the cell, $\delta(z)$ is the Dirac delta function, and $H(z)$ is the Heaviside step function. The first term in the numerator is due to the autocrine ligands, which are recaptured by the same cell from which they were released. The second term in the numerator is due to the paracrine ligands. The fraction of autocrine ligands, $P_{auto}$, is given by

$$P_{auto} = \frac{\tilde{\kappa}}{1 + \tilde{\kappa} + \tilde{k}_b^{1/2}} \qquad 2.9$$

The second term in the denominator is due to the paracrine ligands, which bind to other cells in the medium. The fraction of such ligands, $P_{para}$, is given by

$$P_{para} = 1 - P_{auto} = \frac{1 + \tilde{k}_b^{1/2}}{1 + \tilde{\kappa} + \tilde{k}_b^{1/2}} \qquad 2.10$$

Both $P_{auto}$ and $P_{para}$ can be expressed in terms of $\tilde{\kappa}$ and $\omega$, which characterize individual cells and medium in which the ligands diffuse:

$$P_{auto} = \frac{\tilde{\kappa}(1+\tilde{\kappa})^{1/2}}{(1+\tilde{\kappa})^{3/2} + (3\omega\tilde{\kappa})^{1/2}} \qquad P_{para} = \frac{(1+\tilde{\kappa})^{1/2} + (3\omega\tilde{\kappa})^{1/2}}{(1+\tilde{\kappa})^{3/2} + (3\omega\tilde{\kappa})^{1/2}}$$

The probability density, $p_{para}(r)$, which characterizes the distribution of the trapping points of paracrine ligands, has the following form:

$$p_{para}(r) = \frac{\tilde{k}_b r \exp\left[-(r/R-1)\tilde{k}_b^{1/2}\right]}{R^2 \left(1 + \tilde{k}_b^{1/2}\right)} H(r-R) \qquad 2.11$$

Based on this, the average distance travelled by paracrine ligands, $\langle r_{para} \rangle$, is given by

$$\langle r_{para} \rangle = R\left(1 + \frac{2 + \tilde{k}_b^{1/2}}{\tilde{k}_b^{1/2} + \tilde{k}_b}\right) \qquad 2.12$$

This distance characterizes the length scale of cell communication by secreted ligands.

The probability densities for the lifetimes of autocrine and paracrine ligands, $\varphi_{auto}(t)$ and $\varphi_{para}(t)$, are given by

$$\varphi_{auto}(t) = \frac{D}{R^2}(1+\tilde{\kappa})\left(1+\tilde{\kappa}+\tilde{k}_b^{1/2}\right)\left[\sqrt{\frac{\theta}{\pi t}} - e^{t/\theta}\operatorname{erfc}\left(\sqrt{t/\theta}\right)\right] e^{-k_b t} \qquad 2.13$$

$$\varphi_{para}(t) = \frac{D}{R^2} \frac{\left(1+\tilde{\kappa}+\tilde{k}_b^{1/2}\right)\tilde{k}_b}{(1+\tilde{\kappa})\left(1+\tilde{k}_b^{1/2}\right)}\left[1 + \tilde{\kappa}e^{t/\theta}\operatorname{erfc}\left(\sqrt{t/\theta}\right)\right] e^{-k_b t} \qquad 2.14$$

where $\theta \equiv R^2 / [D(1+\tilde{\kappa})^2]$. From these distribution functions one can find the corresponding average lifetimes of autocrine and paracrine ligands, $\langle t_{auto} \rangle$ and $\langle t_{para} \rangle$:

$$\langle t_{auto} \rangle = \frac{R^2}{2D\tilde{k}_b^{1/2}\left(1+\tilde{\kappa}+\tilde{k}_b^{1/2}\right)} \qquad 2.15$$

$$\langle t_{para} \rangle = \frac{R^2\left[2\left(1+\tilde{k}_b^{1/2}\right)^2 + \tilde{\kappa}\left(2+\tilde{k}_b^{1/2}\right)\right]}{2D\tilde{k}_b\left(1+\tilde{k}_b^{1/2}\right)\left(1+\tilde{\kappa}+\tilde{k}_b^{1/2}\right)} \qquad 2.16$$

As might be expected, $\langle t_{para} \rangle$ is always larger than $\langle t_{auto} \rangle$. In the next two sections, we derive these results and demonstrate their application to the analysis of IFN$\beta$-mediated autocrine signaling in cultures of dendritic cells.

### DERIVATIONS

Consider a ligand released from the surface of a cell located at the origin at $t=0$. To describe the fate of this ligand one has to solve the diffusion equation with partially absorbing boundary conditions on surfaces of randomly located cells and then to average the result over cell configurations. Effective medium approximation allows us to convert this unsolvable problem into a solvable one. This approximation replaces the non-uniform medium by an effective uniform medium (see Fig. 1) in which ligand binding is described by the volumetric rate constant $k_b$, Eq. 2.4.

The probability density of finding the ligand at point **r** at time $t$ is given by the propagator $g(r,t)$ which depends only on the distance $r=|\mathbf{r}|$ because the problem is spherically symmetric. The propagator for this problem satisfies

$$\frac{\partial g}{\partial t} = \frac{D}{r^2}\frac{\partial}{\partial r}\left(r^2 \frac{\partial g}{\partial r}\right) - k_b g, \quad r > R \qquad 3.1$$

with the initial condition

$$g(r,0) = \frac{1}{4\pi R^2}\delta(r-R) \qquad 3.2$$

and the boundary condition on the surface of the "parent" cell located at the origin:

$$D\frac{\partial g(r,t)}{\partial r}\bigg|_{r=R} = \kappa g(R,t) \qquad 3.3$$

Solving this problem, one can find the Laplace transform of $g(r,t)$:

$$\hat{g}(r,s) = \int_0^\infty g(r,t)e^{-st}dt = \frac{\exp\left(-(r-R)\sqrt{(s+k_b)/D}\right)H(r-R)}{4\pi Dr\left[1+\tilde{\kappa}+\sqrt{\tilde{k}_b(1+s/k_b)}\right]} \qquad 3.4$$

where $s$ is the parameter of the Laplace transform. In the rest of this section we use this result to derive the expressions in Eqs. 2.6-2.16.

## A. Ligand lifetime

The survival probability of the ligand before its first binding, $S(t)$, is given by

$$S(t) = 4\pi \int_R^\infty r^2 g(r,t) dr \qquad 3.5$$

Its Laplace transform, $\hat{S}(s)$, can be found using the Laplace transform of the propagator in Eq. 3.4

$$\hat{S}(s) = \int_0^\infty S(t) e^{-st} dt = \frac{1 + \sqrt{\tilde{k}_b (1 + s/k_b)}}{(s + k_b)\left[1 + \tilde{\kappa} + \sqrt{\tilde{k}_b (1 + s/k_b)}\right]} \qquad 3.6$$

Analytical inversion of this transform leads to the result in Eq. 2.6. The average lifetime of the ligand before the first binding, $\langle t \rangle$, is defined by

$$\langle t \rangle = \int_0^\infty t \varphi(t) dt = \hat{S}(0) \qquad 3.7$$

where $\varphi(t) \equiv -dS(t)/dt$ is the probability density for the ligand lifetime. Using Eq. 3.6 we obtain the result in Eq. 2.7.

## B. Distribution of ligand trapping points

The ligand can be trapped either by the parent cell or by one of the cells in the bulk. The probability to be trapped by the parent cell between $t$ and $t+dt$ is $4\pi R^2 \kappa g(R,t) dt$. At the same time, the probability to be trapped at distance between $r$ and $r+dr$, where $r > R$, is $4\pi r^2 k_b g(r,t) dt dr$. Integrating both of these probabilities with respect to time, we get the two marginal probabilities, which lead to the following expression for the probability density of the ligand trapping points:

$$p(r) = \delta(r - R) 4\pi R^2 \kappa \hat{g}(R,0) + 4\pi r^2 k_b \hat{g}(r,0) \qquad 3.8$$

Substituting the expression for the Laplace transform of the propagator, Eq. 3.4, we obtain the result in Eq. 2.8. One can check that $p(r)$ is normalized to unity:

$$\int_R^\infty p(r) dr = 1 \qquad 3.9$$

The first term in Eq. 2.8 is due to autocrine ligands, which bind to the parent cell. The second term is due to paracrine trajectories, which lead to binding by other cells in the medium.

The fractions of the autocrine and paracrine trajectories, $P_{auto}$ and $P_{para}$, are given by

$$P_{auto} = 4\pi R^2 \kappa \int_0^\infty g(R,t) dt = 4\pi R^2 \kappa \hat{g}(R,0) \qquad 3.10$$

$$P_{para} = 4\pi k_b \int_0^\infty \int_0^\infty r^2 g(r,t) dt dr = 4\pi k_b \int_R^\infty r^2 \hat{g}(r,0) dr \qquad 3.11$$

This leads to the expressions in Eqs. 2.9 and 2.10; clearly, $P_{auto} + P_{para} = 1$.

Using $P_{para}$, we introduce the conditional probability density of the trapping points for paracrine trajectories, $p_{para}(r)$:

$$p_{para}(r) = \frac{1}{P_{para}} 4\pi r^2 k_b \hat{g}(r,0) \qquad 3.12$$

Combining this with the Laplace transform of the propagator, Eq. 3.4, we obtain the expression for $p_{para}(r)$ in Eq. 2.11. The first moment of this probability density, gives the average trapping distance for the paracrine ligands, $\langle r_{para} \rangle$, see Eq. 2.12.

## C. Distribution of the lifetimes for autocrine and paracrine trajectories

The probability densities of the lifetimes for autocrine and paracrine trajectories, denoted by $\varphi_{auto}(t)$ and $\varphi_{para}(t)$, are introduced as follows. The fraction of trajectories recaptured by the parent cell is given by $P_{auto}$. This probabilty has contributions from autocrine binding events at all times, from $t=0$ to $t=\infty$, see Eq. 3.10. By definition, $\varphi_{auto}(t)dt$ is the fraction of $P_{auto}$ which is contributed by trajectories/ligands recaptured by between $t$ and $t+dt$. Since the probability to be recaptured by the parent cell between $t$ and $t+dt$ is $4\pi R^2 \kappa g(R,t)dt$, the probability density $\varphi_{auto}(t)$ can be written as:

$$\varphi_{auto}(t) = \frac{1}{P_{auto}} 4\pi R^2 \kappa g(R,t) \qquad 3.13$$

Similarly, the probability density for the binding times in the bulk can be found as:

$$\varphi_{para}(t) = \frac{1}{P_{para}} 4\pi k_b \int_R^\infty r^2 g(r,t) dr \qquad 3.14$$

The Laplace transform of $\varphi_{auto}(t)$ can be found using the Laplace transform of the propagator in Eq. 3.4 and the expression for $P_{auto}$ in Eq. 3.10:

$$\hat{\varphi}_{auto}(s) = \frac{1+\tilde{\kappa}+\tilde{k}_b^{1/2}}{1+\tilde{\kappa}+\sqrt{\tilde{k}_b(1+s/k_b)}} \qquad 3.15$$

Inversion of this transform leads to the expression in Eq. 2.13. The average lifetime of an autocrine ligand, $\langle t_{auto} \rangle$, can be found as follows:

$$\langle t_{auto} \rangle = \int_0^\infty t \varphi_{auto}(t) dt = -\frac{d\hat{\varphi}_{auto}(s)}{ds}\bigg|_{s=0} \qquad 3.16$$

The leads to the expression in Eq. 2.15. A similar sequence of steps leads to the Laplace transform of $\varphi_{para}(t)$:

$$\hat{\varphi}_{para}(s) = \frac{\left(1+\tilde{\kappa}+\tilde{k}_b^{1/2}\right)\left[1+\sqrt{\tilde{k}_b(1+s/k_b)}\right]}{\left(1+\tilde{k}_b^{1/2}\right)(1+s/k_b)\left[1+\tilde{\kappa}+\sqrt{\tilde{k}_b(1+s/k_b)}\right]} \qquad 3.17$$

The inversion of this transform yields the result in Eq. 2.14. Using $\hat{\varphi}_{para}(s)$ we can find the average lifetime of the paracrine trajectories, $\langle t_{para} \rangle = -d\hat{\varphi}_{para}(s)/ds|_{s=0}$, which leads to the expression in Eq. 2.16.

Based on the definitions of $\varphi(t)$, $\varphi_{auto}(t)$, and $\varphi_{para}(t)$, one can see that these probability densities satisfy

$$\varphi(t) = P_{auto}\varphi_{auto}(t) + P_{para}\varphi_{para}(t) \qquad 3.18$$

As a consequence, the average lifetime of secreted ligand, $\langle t \rangle$, is the weighted sum of the average lifetimes of autocrine and paracrine ligands, $\langle t_{auto} \rangle$ and $\langle t_{para} \rangle$:

$$\langle t \rangle = P_{auto}\langle t_{auto} \rangle + P_{para}\langle t_{para} \rangle \qquad 3.19$$

### APPLICATION TO IFN$\beta$ SIGNALING IN DENDRITIC CELLS

We have used these general results to analyze the spatial and temporal ranges of secreted IFNβ molecules in experiments on early responses of cultured human dendritic cells to viral infection (26). In response to viral infection, dendritic cells begin to secrete IFNβ. Once captured by ligand-specific cell surface receptors, IFNβ can induce (after a delay) the secretion of IFNβ and IFNα. To know whether or not the secreted IFNβ will be recaptured by the secreting cell, and to determine the spatial and temporal ranges of IFNβ ligands, we collected values for the molecular, cellular, and physical parameters in this system (see Table 1). Using these parameters we have computed the distribution functions for the trapping distances and the lifetime or autocrine and paracrine ligands, Figure 2. Furthermore, we found that the system operates in the regime of weak binding $\tilde{\kappa} \ll 1$ and small volume fraction of the cells, $R^3 c \ll 1$.

In this regime, the expressions above greatly simplify and reduce to:

$$P_a = k_{on} N_R / (4\pi D R) \qquad 4.1$$
$$<t> = 1/(c k_{on} N_R) \qquad 4.2$$
$$<r> = 2\sqrt{D/(c k_{on} N_R)} \qquad 4.3$$

Using these simple formulas we predict that dendritic cells recapture 2.4% of secreted ligands, that their characteristic travel length is 6 cell-to-cell distances, and their characteristic lifetime in the medium is 20 minutes. Note that the characteristic time and length scales are independent of the cell size. Furthermore, since the system operates in the regime of slow binding, the characteristic time scale is independent of the ligand diffusivity. Clearly, the characteristic ligand trapping distance greatly exceeds both the cell size and the cell-cell distance. This can be considered as an *a posteriori* justification of our effective medium approach to the problem and shows that the analytical approach developed in this paper is perfectly suited for analyzing autocrine signaling in experiments with cultured cells.

### CONCLUDING REMARKS

Based on the effective medium approach, we developed a formalism for analyzing the spatial and temporal ranges of autocrine signaling in cultures of suspended cells. In contrast to the analysis of experiments with adherent cells, which relied on the boundary homogenization approach (19,23,10), our analysis in this paper in based on the homogenization of the heterogeneous three-dimensional medium. The differences between autocrine signaling in the two experimental formats are clearly seen in the dependence of the statistical properties of ligand trajectories on the original parameters of the problems. For instance, one of the key parameters in experiments with adherent cells is the height of the liquid medium. Our previous work has shown that these experiments frequently operate in the regime where the height of the medium can be considered infinite (19,24,10). In this regime, both the average lifetimes of ligands and the average trapping distances are very large, and the kinetics of ligand removal from the medium is stongly nonexponential. This regime does not appear in the experiments with suspended cells. Indeed, in the first format, the ligand spends a lot of time in the medium free from traps and is trapped only at the boundary. In contrast, in the second format, the effective trapping rate in the medium is nonzero in all regions of space.

Our results provide the basis for the development of more complex models of autocrine signaling. Using the statistical properties of individual ligand trajectories derived in this paper, it is possible to analyze the kinetics of ligand accumulation in the medium. This can be most conveniently done using the integral equations, which contain the ligand survival probabilities as their kernels (10). With the model for the extracellular ligand concentrations at hand, it should be possible to link the ligand and receptor part of the problem to the dynamics of intracellular signaling. Recently, such a multimodel approach has been used to analyze the dynamics of autocrine signaling in the EGFR and TNF$\alpha$ systems (4,32,33). Implementing this program for experiments with cultured dendritic cells will rely on the results derived in this paper and on a number of recently published models JAK/STAT signaling pathway stimulated by IFN$\beta$ (34,35).

## ACKNOWLEDGMENTS

This study was supported by the NIH contract HHSN266200500021C ADB No. N01-AI-50021 and the Intramural Research Program of the NIH, Center for Information Technology.

## REFERENCES

1. Kesari, S., and C.D. Stiles. 2006. The bad seed: PDGF receptors link adult neural progenitors to glioma stem cells. *Neuron* 51(2):151-153.
2. Freeman, M. 2000. Feedback control of intercellular signalling in development. *Nature* 408(6810):313-319.
3. Vermeer, P.D., L.A. Einwalter, T.O. Moninger, T. Rokhlina, J.A. Kern, J. Zabner, andM.J. Welsh. 2003. Segregation of receptor and ligand regulates activation of epithelial growth factor receptor. *Nature* 422(6929):322-326.
4. Shvartsman, S.Y., M.P. Hagan, A. Yacoub, P. Dent, H.S. Wiley, andD.A. Lauffenburger. 2002. Autocrine loops with positive feedback enable context-dependent cell signaling. *Am J Physiol Cell Physiol* 282:C545-C559.
5. DeWitt, A., J. Dong, H. Wiley, andD. Lauffenburger. 2001. Quantitative analysis of the EGF receptor autocrine system reveals cryptic regulation of cell response by ligand capture. *J Cell Sci* 114:2301-2313.

6. Dong, J.Y., L.K. Opresko, P.J. Dempsey, D.A. Lauffenburger, R.J. Coffey, and H.S. Wiley. 1999. Metalloprotease-mediated ligand release regulates autocrine signaling through the epidermal growth factor receptor. *Proc Natl Acad Sci* 96(11):6235-6240.
7. Janes, K.A., S. Gaudet, J.G. Albeck, U.B. Nielsen, D.A. Lauffenburger, and P.K. Sorger. 2006. The response of human epithelial cells to TNF involves an inducible autocrine cascade. *Cell* 124(6):1225-1239.
8. Maheshwari, G., H.S. Wiley, and D.A. Lauffenburger. 2001. Autocrine epidermal growth factor signaling stimulates directionally persistent mammary epithelial cell migration. *J Cell Biol* 155(7):1123-1128.
9. Wiley, H.S., S.Y. Shvartsman, and D.A. Lauffenburger. 2003. Computational modeling of the EGF-receptor system: a paradigm for systems biology. *Trends Cell Biol* 13:43-50.
10. Monine, M.I., A.M. Berezhkovskii, E.J. Joslin, H.S. Wiley, D.A. Lauffenburger, and S.Y. Shvartsman. 2005. Ligand accumulation in autocrine cell cultures. *Biophys J* 88(4):2384-2390.
11. Oehrtman, G.T., H.S. Wiley, and D.A. Lauffenburger. 1998. Escape of autocrine ligands into extracellular medium: Experimental test of theoretical model predictions. *Biotechnol Bioeng* 57(5):571-582.
12. Griffith, L.G., and M.A. Swartz. 2006. Capturing complex 3D tissue physiology in vitro. *Nat Rev Mol Cell Biol* 7 (3):211-224.
13. Nakashiro, K., S. Hara, Y. Shinohara, M. Oyasu, H. Kawamata, S. Shintani, H. Hamakawa, and R. Oyasu. 2004. Phenotypic switch from paracrine to autocrine role of hepatocyte growth factor in an androgen-independent human prostatic carcinoma cell line. *Am J Pathol* 165(2):533-540.
14. Shaw, K.R., C.N. Wrobel, and J.S. Brugge. 2004. Use of three-dimensional basement membrane cultures to model oncogene-induced changes in mammary epithelial morphogenesis. *J Mammary Gland Biol Neoplasia* 9(4):297-310.
15. Wrobel, C.N., J. Debnath, E. Lin, S. Beausoleil, M.F. Roussel, and J.S. Brugge. 2004. Autocrine CSF-1R activation promotes Src-dependent disruption of mammary epithelial architecture. *J Cell Biol* 165(2):263-273.
16. Zhu, T., B. Starling-Emerald, X. Zhang, K.O. Lee, P.D. Gluckman, H.C. Mertani, and P.E. Lobie. 2005. Oncogenic transformation of human mammary epithelial cells by autocrine human growth hormone. *Cancer Res* 65(1):317-324.
17. Loewer, A., and G. Lahav. 2006. Cellular conference call: external feedback affects cell-fate decisions. *Cell* 124(6):1128-1130.
18. Lauffenburger, D.A., K.E. Forsten, B. Will, and H.S. Wiley. 1995. Molecular/cell engineering approach to autocrine ligand control of cell function. *Ann Biomed Eng* 23:208-215.
19. Batsilas, L., A.M. Berezhkovskii, and S.Y. Shvartsman. 2003. Stochastic model of autocrine and paracrine signals in cell culture assays. *Biophys J* 85(6):3659-3665.
20. Muratov, C.B., and S.Y. Shvartsman. 2004. Signal propagation and failure in discrete autocrine relays. *Phys Rev Lett* 93(11):118101.
21. Pribyl, M., C.B. Muratov, and S.Y. Shvartsman. 2003a. Discrete models of autocrine signaling in epithelial layers. *Biophys J* 84(6):3624-3635.
22. Pribyl, M., C.B. Muratov, and S.Y. Shvartsman. 2003b. Long-range signal transmission in autocrine relays. *Biophys J* 84:883-896.

23. Berezhkovskii, A.M., Y.M. Makhnovskii, M. Monine, V.Y. Zitserman, and S.Y. Shvartsman. 2004a. Boundary homogenization for trapping by patchy surfaces. *J Chem Phys* 121(22):11390-11394.
24. Berezhkovskii, A.M., L. Batsilas, and S.Y. Shvartsman. 2004b. Ligand trapping in epithelial layers and cell cultures. *Biophys Chem* 107:221-227.
25. Forsten, K.E., and D.A. Lauffenburger. 1994. The Role of Low-Affinity Interleukin-2 Receptors in Autocrine Ligand-Binding - Alternative Mechanisms For Enhanced Binding Effect. *Molecular Immunology* 31:739-751.
26. Fernandez-Sesma, A., S. Marukian, B.J. Ebersole, D. Kaminski, M.S. Park, T. Yuen, S.C. Sealfon, A. Garcia-Sastre, and T.M. Moran. 2006. Influenza Virus Evades Innate and Adaptive Immunity via the NS1 Protein. *J. Virol.* 80(13):6295-6304.
27. Lauffenburger, D.A., and J.J. Linderman. 1993. Receptors: Models for Binding, Trafficking, and Signaling. Oxford University Press, New York.
28. Shvartsman, S.Y., H.S. Wiley, W.M. Deen, and D.A. Lauffenburger. 2001. Spatial range of autocrine signaling: modeling and computational analysis. *Biophys J* 81(4):1854-1867.
29. Shoup, D., and A. Szabo. 1982. Role of Diffusion in Ligand-Binding to Macromolecules and Cell- Bound Receptors. *Biophys J* 40(1):33-39.
30. Rice, S.A. 1985. Diffusion-limited reactions. Elsevier, Amsterdam; New York.
31. Abramowitz, M., and I.A. Stegun. 1964. Handbook of mathematical tables with formulas, graphs, and mathematical tables. Dover Pubns, Washington.
32. Cheong, R., A. Bergmann, S.L. Werner, J. Regal, A. Hoffmann, and A. Levchenko. 2006. Transient IkappaB kinase activity mediates temporal NF-kappaB dynamics in response to a wide range of tumor necrosis factor-alpha doses. *J Biol Chem* 281(5):2945-2950.
33. Maly, I.V., H.S. Wiley, and D.A. Lauffenburger. 2004. Self-organization of polarized cell signaling via autocrine circuits: computational model analysis. *Biophys J* 86(1):10-22.
34. Singh, A., A. Jayaraman, and J. Hahn. 2006. Modeling regulatory mechanisms in IL-6 signal transduction in hepatocytes. *Biotechnol Bioeng* 95(5):850-862.
35. Swameye, I., T.G. Muller, J. Timmer, O. Sandra, and U. Klingmuller. 2003. Identification of nucleocytoplasmic cycling as a remote sensor in cellular signaling by databased modeling. *Proc Natl Acad Sci* 100(3):1028-1033.

TABLE 1 Model parameters

| Parameter | Value |
|---|---|
| $R$ | $25\,\mu m$ |
| $D$ | $10^2\,\mu m^2 s^{-1}$ |
| $c$ | $10^{-6}\,cells/\mu m^3$ |
| $N_R$ | $5\times 10^4\,/cell$ |
| $k_{on}$ | $10^7\,M^{-1}s^{-1}$ |
| $k_{off}$ | $10^{-3}\,s^{-1}$ |
| $k_b$ | $5.1\times 10^{-3}$ |
| $\tilde{\kappa}$ | $2.6\times 10^{-2}$ |

# FIGURE LEGENDS

Figure 1:

Effective medium approximation. The three-dimensional suspension of partially absorbing cells (A) is approximated by an effective medium which is characterized by reaction rate constant $k_b$. (B) Two typical trajectories: autocrine and paracrine.

Figure 2:

(A) Distribution of the ligand trapping points, computed for parameters corresponding to experiments with cultured dendritic cells. Plain line: standard parameters $\tilde{k}_b = 5.1 \times 10^{-3}$ ($\langle r_{para} \rangle / R \approx 28 \approx 6$ cell-to-cell distances), dashed line: $\tilde{k}_b$ five times decreased ($\langle r_{para} \rangle / R \approx 13 \approx 3$ cell-to-cell distances), dotted line: $\tilde{k}_b$ five times increased ($\langle r_{para} \rangle / R \approx 63 \approx 13$ cell-to-cell distances). Note that the maximum of the probability density is shifted toward the parent cell origin as $\tilde{k}_b$ - or equivalently the cell concentration $c$ - increases.

(B) Densities of the conditional survival probabilities for autocrine trajectories (solid line) and paracrine trajectories (dashed line). Parameters used to generate these plots are given in Table 1. Inside caption: value of the mean characteristic lifetimes, computed as the means of the corresponding distribution functions, see text for details.

**FIGURE 1**

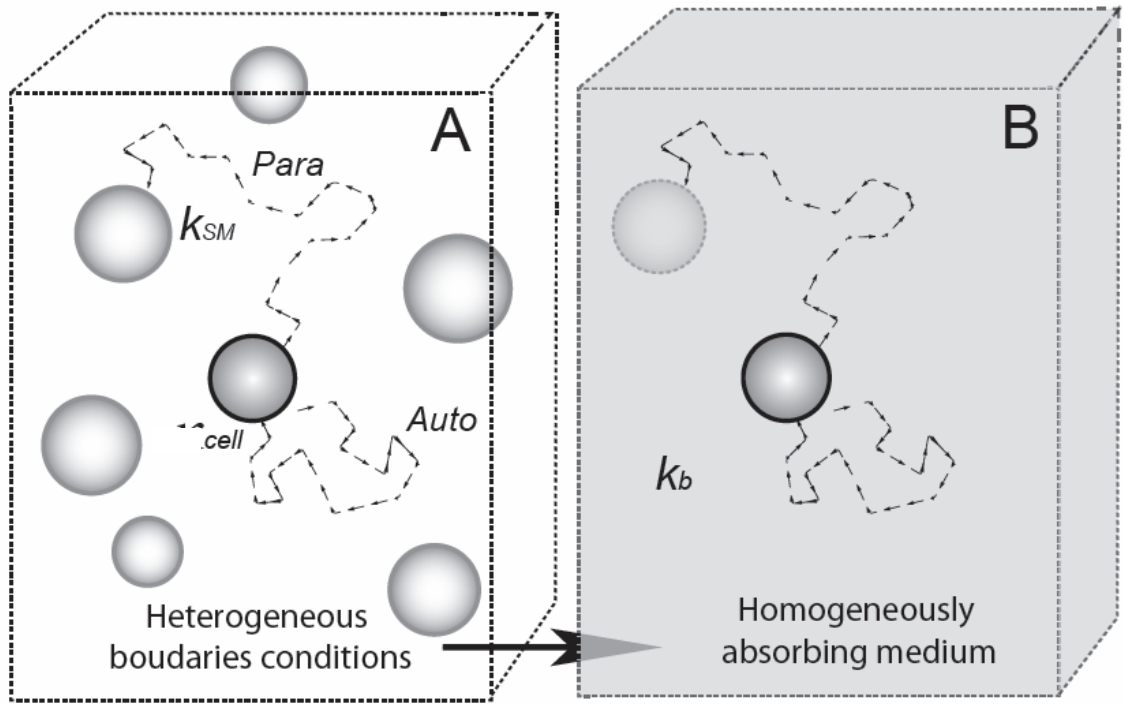

**FIGURE 2**

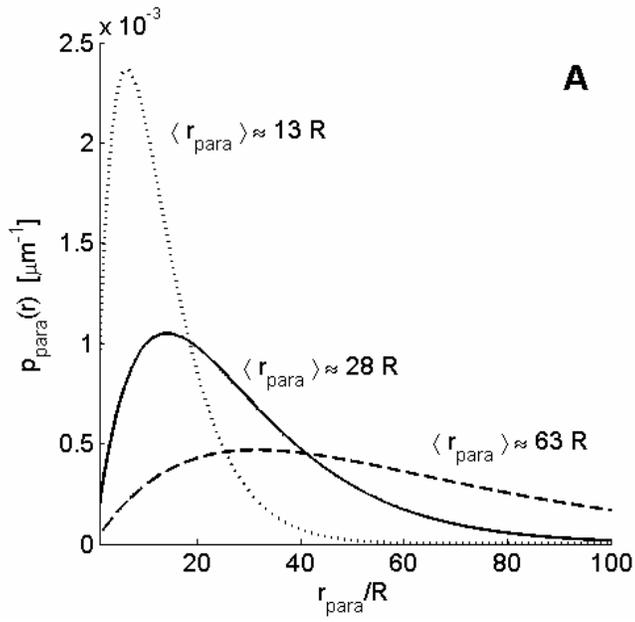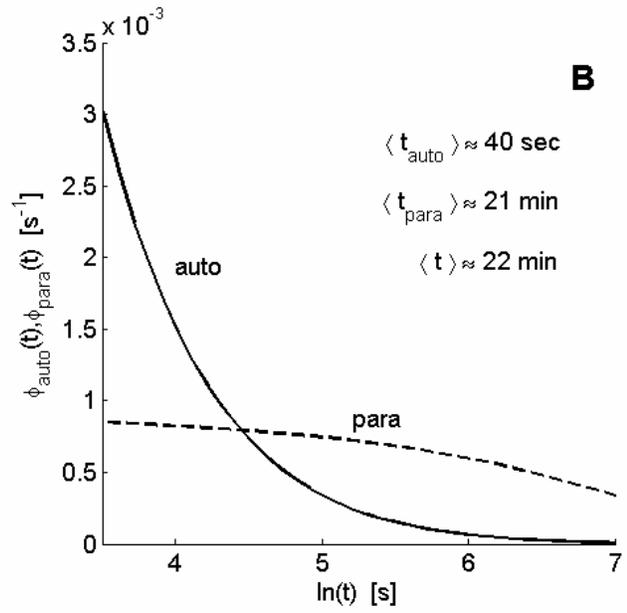